\documentclass[11pt]{article}
\usepackage{epsfig}
\def\be{\begin{equation}}
\def\ee{\end{equation}}
\def\bc{\begin{center}}
\def\ec{\end{center}}
\def\bea{\begin{eqnarray}}
\def\eea{\end{eqnarray}}

\def\nn{\nonumber}

\def\ov{\overline}
\def\hlf{\frac{1}{2}}

\def\as{\alpha_s}
\def\at{\alpha_t}
\def\ab{\alpha_b}
\def\sq2{\sqrt{2}}
\def\ths{\theta_{\tilde{t}}}
\def\thz{\theta_{\tilde{b}}}
\def\csenza{c_{2 \theta_b}^{\,2}}
\def\s2t{s_{2\theta_b}}
\def\c2t{c_{2\theta_b}}
\def\mgl{m_{\tilde{g}}}
\def\msqu{m_{\tilde{b}_1}^2}
\def\msqd{m_{\tilde{b}_2}^2}
\def\msqut{m_{\tilde{t}_1}^2}
\def\msqdt{m_{\tilde{t}_2}^2}
\def\mbu{m_{\tilde{b}_1}}
\def\mbd{m_{\tilde{b}_2}}

\def\sb{s_\beta}
\def\cb{c_\beta}
\def\tb{\tan \beta}

\newcommand{\smallz}{{\scriptscriptstyle Z}} %
\newcommand{\mz}{m_\smallz}
\newcommand{\ma}{m_{\scriptscriptstyle A}}

\newcommand{\mh}{m_h}
\newcommand{\mH}{m_{\scriptscriptstyle H}}
\def\li2{{\rm Li}_2}
\def\msquq{m_{\tilde{b}_1}^4}
\def\msqdq{m_{\tilde{b}_2}^4}
\def\diff{\msqu-\msqd}

\def\mylg{\ln\frac{\mgl^2}{Q^2}}

\def\myltu{\ln\frac{\msqu}{Q^2}}
\def\myltd{\ln\frac{\msqd}{Q^2}}

\def\myltuq{\ln^2\frac{\msqu}{Q^2}}
\def\myltdq{\ln^2\frac{\msqd}{Q^2}}

\catcode`@=11
\def\marginnote#1{}
\newcount\hour
\newcount\minute
\newtoks\amorpm
\hour=\time\divide\hour by60
\minute=\time{\multiply\hour by60 \global\advance\minute by-\hour}
\edef\standardtime{{\ifnum\hour<12 \global\amorpm={am}%
        \else\global\amorpm={pm}\advance\hour by-12 \fi
        \ifnum\hour=0 \hour=12 \fi
        \number\hour:\ifnum\minute<10 0\fi\number\minute\the\amorpm}}
\edef\militarytime{\number\hour:\ifnum\minute<10 0\fi\number\minute}
\def\draftlabel#1{{\@bsphack\if@filesw {\let\thepage\relax
   \xdef\@gtempa{\write\@auxout{\string
      \newlabel{#1}{{\@currentlabel}{\thepage}}}}}\@gtempa
   \if@nobreak \ifvmode\nobreak\fi\fi\fi\@esphack}
        \gdef\@eqnlabel{#1}}
\def\@eqnlabel{}
\def\@vacuum{}
\def\draftmarginnote#1{\marginpar{\raggedright\scriptsize\tt#1}}
\def\draft{\oddsidemargin 0.0truein
        \def\@oddfoot{\sl preliminary draft \hfil
        \rm\thepage\hfil\sl\today\quad\militarytime}
        \let\@evenfoot\@oddfoot \overfullrule 3pt
        \let\label=\draftlabel
        \let\marginnote=\draftmarginnote
   \def\@eqnnum{(\theequation)\rlap{\kern\marginparsep\tt\@eqnlabel}%
\global\let\@eqnlabel\@vacuum}  }
\catcode`@=12
\begin{document}
\thispagestyle{empty}
\begin{center}
\hfill{DFPD-02/TH/05} \\
\hfill{RM3-TH/02-06}  \\
\hfill{ROME1-1332/02} \\
\vspace{1.7cm}
\bc
{\LARGE\bf On the  two--loop sbottom corrections to the}
\ec
\bc
{\LARGE\bf neutral Higgs boson masses in the MSSM}
\ec
\vspace{1.4cm}
{\Large \sc A.~Brignole~$^{a}$, G.~Degrassi~$^{b}$,
P.~Slavich~$^{c}$ and F.~Zwirner~$^{d}$}\\

\vspace{1.2cm}

${}^a$
{\em 
Dipartimento di Fisica `G.~Galilei', Universit\`a di Padova and
\\ 
INFN, Sezione di Padova, Via Marzolo~8, I-35131 Padua, Italy
}
\vspace{.3cm}

${}^b$
{\em 
Dipartimento di Fisica, Universit\`a di Roma III and 
\\
INFN, Sezione di Roma III, Via della Vasca Navale~84, I-00146 Rome, Italy 
}
\vspace{.3cm}

${}^c$
{\em 
Physikalisches Institut der Universit\"at Bonn,
\\
Nussallee 12, D-53115 Bonn, Germany}
\vspace{.3cm}

${}^d$
{\em 
Dipartimento di Fisica, Universit\`a di Roma `La Sapienza' and
\\
INFN, Sezione di Roma, P.le Aldo Moro~2, I-00185 Rome, Italy}

\end{center}

\vspace{0.8cm}

\centerline{\bf Abstract}
\vspace{2 mm}
\begin{quote} \small
  We compute the ${\cal O} ( \ab \as)$ two--loop corrections to the
  neutral Higgs boson masses in the Minimal Supersymmetric Standard
  Model, using the effective potential approach. Such corrections can
  be important in the region of parameter space corresponding to $\tb
  \gg 1$ and sizeable $\mu$. In spite of the formal analogy with the
  ${\cal O} ( \at \as)$ corrections, there are important differences,
  since the dominant effects are controlled by the sbottom--Higgs
  scalar couplings. We propose a convenient renormalization scheme
  that avoids unphysically large threshold effects associated with the
  bottom mass, and absorbs the bulk of the ${\cal O} (\ab \as + \ab
  \at)$ corrections into the one--loop expression. We give general
  explicit formulae for the ${\cal O} (\ab \as)$ corrections to the
  neutral Higgs boson mass matrix. We also discuss the importance of
  the ${\cal O} (\ab^2)$ corrections and derive a formula for their
  contribution to $m_h$ in a simple limiting case.
\end{quote}
\vfill
\newpage
\setcounter{equation}{0}
\setcounter{footnote}{0}
\vskip2truecm
\section{Introduction}

The existence of a light CP--even neutral Higgs boson is a crucial
prediction of the Minimal Supersymmetric extension of the Standard
Model, or MSSM, and has been one of the most active areas of
theoretical investigations in the last decade. At the tree level, the
masses of the neutral CP--even Higgs bosons of the MSSM can be
computed in terms of three input parameters: the mass $\ma$ of the
neutral CP--odd particle, the mass $\mz$ of the weak neutral gauge
boson, and the ratio of Higgs vacuum expectation values $\tb \equiv
v_2/v_1$ (for a review and references, see e.g. \cite{hunter}). For
$\tb \ll m_t/m_b$, the dominant one--loop corrections are the ${\cal
O}(\at)$ ones, where $\at \equiv h_t^2 / (4 \pi)$ and $h_t$ is the
superpotential top coupling.  Such coupling controls both the
top--Higgs Yukawa couplings and a number of cubic and quartic
stop--Higgs scalar couplings, and leads to significant contributions
from both top and stop loops \cite{old}. The ${\cal O}(\ab)$ one--loop
corrections associated with the superpotential bottom coupling $h_b$,
where $\ab \equiv h_b^2 / (4 \pi)$, can be numerically non--negligible
only for $\tb \gg 1$ and sizeable values of the $\mu$ parameter. At the
classical level $h_b / h_t = (m_b / m_t) \tb$, thus we need $\tb \gg
1$ to have $\ab \sim \at$ in spite of $m_b \ll m_t$. Moreover, and in
contrast with the top--stop case, numerically relevant contributions
can only come from sbottom loops: those coming from bottom loops are
always suppressed by the small value of the bottom mass. A sizeable
value of $\mu$ is then required to have sizeable sbottom--Higgs scalar
interactions in the large $\tb$ limit.

We are now at the stage where the most important genuine two--loop
corrections are being evaluated: general results have been obtained
both for the ${\cal O} (\at \as)$ \cite{hh,atas,dsz} and for the
${\cal O} (\at^2)$ \cite{hh,atsq,bdsz} corrections. In this paper we
move one step further, computing the ${\cal O} (\ab \as)$ corrections
and discussing the ${\cal O} (\ab^2)$ and ${\cal O} (\at \ab)$
ones. For convenience, we evaluate two--loop effects directly in the
physically relevant limit of large $\tb$:
\be
\label{ltb}
v_1 \rightarrow 0 \, , 
\;\;\;\;\;
v_2 \rightarrow v \equiv (\sqrt{2}\, G_\mu)^{-1/2} \, ,
\ee
where $G_\mu$ is the Fermi constant.  As a result, we obtain extremely
compact analytical formulae. Keeping $v_1 \neq 0$ would only generate
more complicated expressions, without adding any relevant information.

The plan of the paper is the following. We first give the analytical
result at ${\cal O} (\ab \as)$ and in the $\ov{\rm DR}$ scheme. We
then identify a convenient renormalization scheme that avoids
unphysically large threshold effects and absorbs the largest ${\cal O}
(\ab \as + \ab \at)$ corrections into the one--loop expressions. In
particular, we discuss how to use the experimental information on the
bottom mass, which receives large threshold corrections \cite{hrs}, to
extract the value of the renormalized coupling $h_b$. We finally
present numerical results for some representative parameter choices,
and conclude with an explicit formula for the ${\cal O} (\ab \as +
\ab^2)$ corrections to $m_h$ in a simple limiting case.

\section{General formulae and $\ov{\rm DR}$ results}

The momentum--independent part of the one--loop ${\cal O} (\ab)$ and
two--loop ${\cal O} (\ab \as)$ corrections to the neutral CP--even
Higgs boson mass matrix can be obtained by taking the second
derivatives of the effective potential~\footnote{The effective
potential for vanishing CP--odd fields was computed in \cite{atsq}. To
make contact with the physical $\ma$, the effective potential should
be computed as a function of both CP--even and CP--odd fields, as in
\cite{dsz}.} at its minimum, or by performing appropriate
substitutions and limits in the ${\cal O} (\at \as)$ results of
\cite{dsz}. In the limit of Eq.~(\ref{ltb}), we find:
\bea
\label{dmb11n}
\hspace{-1cm}
\left(\Delta {\cal  M}^2_S\right)_{11}^{\rm eff} 
& = & \frac12 \, h_b^2 \, \s2t^2 \left[ A_b^2 \left( F^{\rm 1\ell} + 
F^{\rm 2\ell} \right) + 2 \, A_b \,  \mgl \,
G^{\rm 2\ell}    \right]  \, , \\
\label{dmb12n}
\hspace{-1cm}
\left(\Delta {\cal  M}^2_S\right)_{12}^{\rm eff} 
& = & \frac12 \, h_b^2 \, \s2t^2 \left[\mu \, A_b \, \left( F^{\rm 1\ell} + 
F^{\rm 2\ell} \right) +\mu \, \mgl \, G^{\rm 2\ell} \right] \, , \\
\label{dmb22n}
\hspace{-1cm}
\left(\Delta {\cal  M}^2_S\right)_{22}^{\rm eff} 
& = &
\frac12\, h_b^2 \, \s2t^2 \, \mu^2\, \left( F^{\rm 1\ell} + F^{\rm 2\ell}
\right) \, .
\eea
Before explaining the meaning of the different symbols, we recall that
an important simplification occurs if we look at the lightest Higgs
eigenvalue, $\mh$, in the limit $\ma \gg \mz$, since in that limit
$\Delta m_h^2$ coincides with $(\Delta {\cal M}^2_S )_{22}^{\rm eff}$.
Our conventions are such that, at the classical level, the top and
bottom quark masses are given by $m_t = h_t v_2 / \sqrt{2}$ and $m_b =
h_b v_1 / \sqrt{2}$, where the Yukawa couplings $(h_t,h_b)$ and the
VEVs $(v_1,v_2)$ are all taken to be real and positive. In addition,
we assume $\mu$ and $A_b$ to be real, but we do not make any
assumption on their sign, whereas we choose the gluino mass $\mgl$ to
be real and positive.  At the classical level, the sbottom mixing
angle $\s2t \equiv \sin 2\theta_{\tilde{b}}$ is given by
\be
s_{2\theta_{b}} = 
\frac{\sqrt{2} \, h_b ( A_b \, v_1 + \mu \, v_2 )}{m_{\tilde{b}_1}^2 
-m_{\tilde{b}_2}^2} \longrightarrow \frac{\sqrt{2} \, h_b \, \mu \,
v }{m_{\tilde{b}_1}^2 - m_{\tilde{b}_2}^2} \, ,
\label{s2b}
\ee
where the arrow denotes the large $\tb$ limit, and $\msqu > \msqd$ are
the two eigenvalues of the sbottom mass matrix. Finally, the
superscripts in the functions $(F,\,G)$ indicate the order of the
loop contribution. At one loop, and in the large $\tb$ limit, the only
relevant function is
\be
\label{G31l}
F^{\rm 1\ell} = \frac{N_c}{16\,\pi^2} \left( 2 -
\frac{\msqu+\msqd}{\msqu-\msqd} \,\ln\frac{\msqu}{ \msqd} \right) \, ,
\ee 
where $N_c = 3$ is a color factor. Notice that $F^{\rm 1\ell}$ is
negative definite.

We first present our results for $F^{\rm 2\ell}$ and $G^{\rm 2\ell}$ in 
the $\ov{\rm DR}$ scheme. In other words, we assume that
the ${\cal O}(\ab)$ one--loop contribution is written entirely in
terms of $\ov{\rm DR}$ parameters (masses and couplings), evaluated at
a certain renormalization scale $Q$.  In units of $g_s^2 \,C_F\,N_c/
(16\,\pi^2)^2$, where $C_F = 4/3$, we find:
\bea
\label{f32l}
\hat{F}^{\rm 2\ell} & = & 
\left( 2 -
\frac{\msqu+\msqd}{\msqu-\msqd}
\,\ln\frac{\msqu}{ \msqd} \right) 
\, \left( 3 + 9 \,c^2_{2 \theta_b} \right)
-\frac{3 + 13\,\csenza}{\diff}\,
\left(\msqu\,\myltu - \msqd\,\myltd\right)\nn \\ 
\hspace{-2cm}&&\nn\\
\hspace{-2cm}&&
+ \,3\,\frac{\msqu +\msqd}{\msqu - \msqd}\,
\left( \myltuq - \myltdq \right)  
+ 4 
- \csenza\, \left[ 
4 - \left(\frac{\msqu+\msqd}{\diff}\right)^2\,\ln^2\frac{\msqu}{\msqd}
\right.\nn\\
\hspace{-2cm}&&\nn\\
\hspace{-2cm}&& \left. 
- 6\,\frac{\msqu+\msqd}{(\diff)^2}\,
\left(\msqu\,\myltu - \msqd\,\myltd\right)\,\ln\frac{\msqu}{\msqd}\right]
- \, \s2t^2\,\left[
\frac{\msqu}{\msqd} + \frac{\msqd}{\msqu}\right.\nn\\
\hspace{-2cm}&&\nn\\
\hspace{-2cm}&& 
\left.
+ 2\,\ln\frac{\msqu\,\msqd}{Q^4}
- \frac{\msquq}{\msqd\,(\diff)}\,\myltu
+ \frac{\msqdq}{\msqu\,(\diff)}\,\myltd\,
\right]\nn\\
\hspace{-2cm}&&\nn\\
\hspace{-2cm}&&
+ \frac{4}{\diff} \left\{
- \frac{\msqd\, \mgl^2 }{\msqu} + 
\frac{\mgl^2\msqd}{\msqu}\,\mylg
- 2\,\mgl^2\,\myltu \right. \nn\\
\hspace{-2cm}&&\nn\\
\hspace{-1.5cm}&& \left.
+ \, \left( 2 \, \mgl^2 -\msqu-\msqd \right) \left[
\,\ln \frac{\mgl^2}{Q^2} \, \myltu - 
\li2 \left( 1- \frac{\msqu}{\mgl^2} \right) \right]
- \left(1 \leftrightarrow 2 \right)\;\right\} \, , \\
\hspace{-2cm}&&\nn\\
\label{f22l}
\hat{G}^{\rm 2\ell} & = &
\frac{4}{\diff} \left[
-4\,\msqu + \left(3\,\msqu + \msqd \right) \,\myltu 
-\, \left( \msqu + \msqd \right)\,\myltu \mylg \right.
\nn  \\
\hspace{-2cm}&&\nn\\
\hspace{-1cm}&&
\hspace{2.5cm}
\left. -\left( 2\,\mgl^2 - \msqu - \msqd \right)\,
\li2 \left(1- \frac{\msqu}{\mgl^2} \right) - \left(
1 \leftrightarrow 2 \right) \right] \, , 
\eea
where the hats on $F$ and $G$ denote ${\rm \ov{DR}}$--quantities, and,
here and in the following, $(1 \leftrightarrow 2)$ denotes the
interchange $\msqu \leftrightarrow \msqd$. Notice that, in our limit,
the bottom quark only contributes through bottom--sbottom--gluino
diagrams.  The above way of presenting the results is convenient for
analysing models that predict, via the MSSM renormalization group
equations, the low--energy $\ov{\rm DR}$ values of the MSSM input
parameters in terms of a more restricted set of parameters, assigned
as boundary conditions at some scale much larger than the weak
scale. One of the parameters, however, is the $\ov{\rm DR}$ coupling
$\hat{h}_b$, which must be connected with the experimental information
on the bottom mass: this issue will be discussed extensively in
Section~4.

\section{A convenient renormalization prescription}

General low--energy analyses of the MSSM parameter space do not
refer to boundary conditions at high scales. These analyses are
usually performed in terms of parameters with a more direct physical
interpretation, such as pole masses and appropriately defined mixing
angles in the squark sector. Such an approach requires modifications 
of our two--loop formulae, Eqs.~(\ref{f32l})--(\ref{f22l}), induced by the
variation of the one--loop parameters when moving from the $\ov{\rm
DR}$ scheme to a different scheme. We recall that, at the one--loop
level, the two VEVs $(v_1,v_2)$ and the mass parameter $\mu$ are not
renormalized by the strong interactions. Therefore, the only
parameters in the Higgs mass matrix that require a one--loop
definition are ($h_b, \, A_b,\, \s2t, \mbu,\,\mbd$), although 
only four of these are independent, because of the relation (\ref{s2b}).
 
The sbottom masses $(\msqu,\msqd)$ in Eq.~(\ref{G31l}) can be
naturally identified with the pole masses. For the generic parameter
$x$, we define the shift from the $\ov{\rm DR}$ value $\hat{x}$ as
$\delta x \equiv \hat{x} - x$. According to this definition, we find
\bea
\delta m_{\tilde{b}_{1}}^2 \equiv
 \Pi_{11}(\msqu) & = &  \frac{g_s^2\,C_F}{16 \pi^2} \,\msqu\,
 \left[ 3\, \ln \frac{\msqu}{Q^2} - 3 -
 c^2_{2 \theta_b} \left( \ln \frac{\msqu}{Q^2} -1 \right) 
-\s2t^2 \frac{\msqd}{\msqu} \left( \ln \frac{\msqd}{Q^2} -1 \right)
 \right. \nn \\
&&\nn\\
& & \left. -6 \frac{\mgl^2}{\msqu} 
- 2 \left( 1 - 2  \frac{\mgl^2}{\msqu} \right) \ln \frac{\mgl^2}{Q^2}
- 2 \left( 1-\frac{\mgl^2}{\msqu} \right)^2
\ln \left| 1-\frac{\msqu}{\mgl^2} \right| \, \right] \, ,
\label{dmb1}
\eea
where $\Pi_{ij}(p^2)$ denotes the real and finite part of the $(ij)$
component of the sbottom self--energy $(i,j=1,2)$, and $\delta\msqd$
is obtained from Eq.~(\ref{dmb1}) by the interchange $\msqu
\leftrightarrow \msqd$.

The most convenient definition of $(h_b,A_b,\s2t)$ is less easily
singled out. To clarify this point, we recall the parallel case of the
${\cal O}(\at \as )$ corrections. In that case, besides the stop pole
masses, the remaining independent parameters are chosen to be
\cite{atas,dsz} a conveniently defined stop mixing angle,
$s_{2\theta_t}$, and the top Yukawa coupling $h_t^{pole}$, as defined
by the top pole mass $M_t$ via the relation $M_t \equiv h_t^{pole} v_2
/ \sqrt{2}$. Then, the stop counterpart of Eq.~(\ref{s2b}) is used to
establish the one--loop definition of $A_t$ in terms of the pole top
and stop masses and of the stop mixing angle. In the case of the
${\cal O}(\ab \as )$ corrections, a similar procedure is not
appropriate since, as can be easily seen from Eq.~(\ref{s2b}), $\s2t$
is independent of $A_b$ in the large $\tb$ limit. A second
complication arises from the large one--loop threshold corrections
\cite{hrs} proportional to $v_2$ that contribute to the pole bottom
mass: for our calculation, the relevant ones are those ${\cal
O}(\as)$, associated with one--loop SQCD diagrams with gluinos and
sbottom quarks on the internal lines. 
As noticed in \cite{eberl}, a definition of $A_b$ in terms
of the pole bottom and sbottom masses through Eq.~(\ref{s2b}) would
produce very large shifts in $A_b$ with respect to its $\ov{\rm DR}$
value, $\delta A_b = {\cal O}(\as \, \mu^2 \, \tan^2 \beta / \mgl)$.
A $\ov{\rm DR}$ definition for the parameters $(h_b, A_b, \s2t)$ would
avoid this problem, but would still suffer from the known fact that it
does not make manifest the decoupling of heavy particles, for example
a heavy gluino.

We then look for definitions of the relevant parameters that
automatically include the decoupling of heavy gluinos, allow to
disentangle the genuine two--loop effects from the large threshold
corrections to the bottom mass, and provide a consistent prescription
for $A_b$ in the large $\tb$ limit. There are two quantities that have
a natural physical interpretation,
\be
\widetilde{X}_b  =  {h_b \, v \over \sqrt{2}} (\cb \, A_b + \sb \,  \mu)
\longrightarrow {h_b \, v \, \mu \over \sqrt{2}} \, ,
\;\;\;\;\;
\widetilde{Y}_b  =   {h_b \over \sqrt{2}} (\sb \, A_b - \cb \, \mu) 
\longrightarrow {h_b \, A_b \over \sqrt{2}} \, ,
\label{XY}
\ee
where the arrows denote as before the large $\tb$ limit.  At the
classical level, $\widetilde{X}_b$ is the off--diagonal term in the
sbottom mass matrix, related to the mixing angle $s_{2\theta_{b}}$ via
Eq.~(\ref{s2b}), and $\widetilde{Y}_b$ is proportional to the
coefficient of the trilinear $(\widetilde{b}_L \widetilde{b}_R^* A)$
interaction, or, equivalently, of the $(\widetilde{b}_1
\widetilde{b}_2^* A)$ interaction.

A suitable definition of the mixing angle $\theta_b$, with the virtue
of being infrared (IR) finite and gauge--independent with respect to
the strong interaction, is~\cite{guasch}:
\be
\label{thetab}
\delta\thz = \hlf \,
\frac{\Pi_{12}(\msqu) + \Pi_{12}(\msqd)}{\msqu-\msqd} \, ,
\ee
where  $\Pi_{12}(p^2)$ turns out to be
independent of $p^2$ in the large $\tb$ limit. Using Eq.~(\ref{s2b}),
the prescription on $\thz$ can be immediately translated into a 
prescription for $\widetilde{X}_b$:
\be
\label{deltaXb}
\delta \widetilde{X}_b = {1 \over 2} \cos 2\theta_b \left[ \Pi_{12}
(\msqu) + \Pi_{12}(\msqd) \right] + \widetilde{X}_b \, 
\frac{\Pi_{11}(\msqu) - \Pi_{22}(\msqd)} {\msqu- \msqd} \, .
\ee
Since, in the large $\tb$ limit, $v$ and $\mu$ are
not renormalized by the strong interactions, the prescription on
$\widetilde{X}_b$ can in turn be translated into a prescription for
$h_b$. Explicitly:
\bea
\delta h_b &=& \frac{g_s^2\,C_F}{16 \pi^2}\, h_b\,\left\{
-4 + 2\, \ln\frac{\mgl^2}{Q^2} 
\phantom{\left(\left(1-\frac{\mgl^2}{\msqu}\right)^2
\ln \left|1 - \frac{\msqu}{\mgl^2}\right|\right)}
\right. \nn\\
&&  
\left.
+ \left[ \frac{2\,\msqu}{\msqu-\msqd}\,
\left( 2\,\ln\frac{\msqu}{\mgl^2} -\left(1-\frac{\mgl^2}{\msqu}\right)^2
\ln \left|1 - \frac{\msqu}{\mgl^2}\right|\right)
+\left( 1 \leftrightarrow 2 \right) \right] \right\} \;.
\label{dhb}
\eea
We stress that our renormalized $h_b$, as defined above, differs at
the one--loop level both from the $\ov{\rm DR}$ quantity $\hat{h}_b$
and from the quantity $h_b^{pole}$ that would be obtained by plugging
the pole bottom mass, $M_b$, into the tree--level formula:
\be
h_b \ne h_b^{pole} \equiv {M_b \sqrt{2} \over v_1} \; .
\ee

Concerning the definition of $A_b$, we observe that the Yukawa
coupling $h_b$ multiplying $A_b$ can be absorbed in a redefinition of
the trilinear soft--breaking term, $\widetilde{A}_b \equiv h_b A_b$.
The shift in $\widetilde{A}_b$ could be defined via a physical
process, e.g. one of the decays $\widetilde{b}_1 \rightarrow
\widetilde{b}_2 \, A$ or $A \rightarrow \widetilde{b}_1 \,
\widetilde{b}_2^*$, but such a definition would suffer from the
problem of infrared (IR) singularities associated with gluon
radiation. To overcome this problem, and given our ignorance of the
MSSM spectrum, we find less restrictive to define
$\delta\widetilde{A}_b$ in terms of the $(\widetilde{b}_1
\widetilde{b}_2^* A)$ proper vertex, at appropriately chosen external
momenta and including suitable wave function corrections, so that the
resulting combination is IR finite and gauge--independent, and gives
rise to an acceptable heavy gluino limit. Denoting the proper vertex
$\tilde{b}_1 \tilde{b}_2^* A$ with $i \Lambda_{12A}(p_1^2, p_2^2,
p_A^2)$, we define~\footnote{This definition is suitable at ${\cal
O}(\as)$. It can be generalized to the case of Yukawa corrections by
specifying a prescription for the $A$ wave function.}:
\bea
\label{deltaab}
\delta\widetilde{A}_b & = & - {i \over \sqrt{2}}\left[
\Lambda_{12A}(\msqu,\msqu,0) + \Lambda_{12A}(\msqd,\msqd,0) \right] \nn\\
&&\nn\\
& & + {1 \over 2}\, \widetilde{A}_b \, 
\frac{ \Pi_{11}(\msqu)+ \Pi_{22}(\msqu) 
- \Pi_{11}(\msqd) - \Pi_{22}(\msqd)} {\msqu- \msqd} ~.
\eea
The above definition can be interpreted as the large $\tb$
limit of a renormalization prescription on $\widetilde{Y}_b$, as
defined in Eq.~(\ref{XY}), since in that limit $\widetilde{Y}_b
\rightarrow \widetilde{A}_b / \sqrt{2}$. Notice the strong resemblance with 
the corresponding renormalization prescription on $\widetilde{X}_b$,
Eq.~(\ref{deltaXb}). At ${\cal O} (\as)$, gauge independence and IR
finiteness follow from the fact that one--loop gluon diagrams satisfy
the identity
\be 
\left[ \Lambda_{12A}(p^2,p^2,0) \right]_g = {i \widetilde{A}_b \over
\sqrt{2}} \frac{\left[ \Pi_{11}(p^2)- \Pi_{22}(p^2) \right]_g }
{\msqu- \msqd} \, ,
\ee
so that the gluon contribution to $\delta\widetilde{A}_b$ can be
written simply as 
\be 
[\delta\widetilde{A}_b]_g = \widetilde{A}_b \frac{\left[
\Pi_{11}(\msqu)- \Pi_{22}(\msqd) \right]_g } {\msqu- \msqd} \, ,
\ee
where the on--shell self--energies $\Pi_{11}(\msqu)$ and
$\Pi_{22}(\msqd)$ are indeed gauge--independent and IR finite.
Writing
\be
\delta\widetilde{A}_b = \delta h_b \, A_b + h_b \, \delta A_b \, ,
\ee
we find
\be 
\delta A_b =
\frac{g_s^2\,C_F}{8 \pi^2} \, \mgl \left\{ 
4- 2\, \ln \frac{\mgl^2}{Q^2} -\left[ \left( 
1 - \frac{\mgl^2}{\msqu} \right) \, 
\ln \left|1 - \frac{\msqu}{\mgl^2} \right|
+ \left( 1 \leftrightarrow 2 \right) \right]
\right\} \, . 
\label{dYb}
\ee

With our one--loop specifications of $h_b$ and $A_b$, Eqs.~(\ref{dhb})
and (\ref{dYb}), the CP--even Higgs boson mass matrix takes again the
form of Eqs.~(\ref{dmb11n})--(\ref{dmb22n}), but the one--loop part
of the corrections must now be evaluated in our renormalization
scheme, and the functions $F^{\rm 2\ell}$ and $G^{\rm 2\ell}$ read
now, in units of $g_s^2 \,C_F\,N_c/(16\,\pi^2)^2$:
\bea 
F^{\rm 2\ell} &=& - (1 + \s2t^2) \left( 2 -
\frac{\msqu+\msqd}{\msqu-\msqd}
\, \ln \frac{\msqu}{\msqd} \right)^2 -
\left( \ln \frac{\msqu}{\msqd} \right)^2 \nn \\
&&\nn\\
&+& 4 -2 \left( {\mgl^2 \over \msqu} +  {\mgl^2 \over \msqd}
\right) + 4 \left( \ln { \msqu \over \mgl^2} + 
\ln {\msqd \over \mgl^2} \right)\nn \\
&&\nn\\
&+& 4 \frac{\msqu+\msqd -2 \,\mgl^2}{ \msqu-\msqd}
\left[ \li2 \left(1- \frac{\msqu}{\mgl^2} \right)
 - \li2 \left(1- \frac{\msqd}{\mgl^2} \right)
- {1 \over 2} \ln\frac{\msqu}{ \msqd} \right] \nn \\
&&\nn\\
&-& \!2 \left[ \left( 1 + \frac{2 \,\msqu}{ \msqu-\msqd}
- \frac{2 \,\msquq}{(\msqu-\msqd)^2}
\ln\frac{\msqu}{ \msqd} \right) \!
\left( 1 -  {\mgl^2 \over \msqu} \right)^2
\ln \left| 1 -  { \msqu \over \mgl^2} \right| 
+ \left( 1 \leftrightarrow 2 \right) \right] \, ,
\nn\\
& & \label{G3bar} \\
& & \nn \\
G^{\rm 2\ell} &=& 
4\, \ln { \msqu \msqd \over \mgl^4} 
+ 4\, \frac{\msqu+\msqd -2 \,\mgl^2}{ \msqu-\msqd} \left[ \li2
  \left(1- \frac{\msqu}{\mgl^2} \right) - \li2 \left(1-
    \frac{\msqd}{\mgl^2} \right) \right]
\nonumber \\
&&\nn\\
& - & \!\!\!  2\,\left( 2 - \frac{\msqu+\msqd}{\msqu-\msqd}
  \,\ln\frac{\msqu}{ \msqd} \right) \left[ \left( 1 - {\mgl^2 \over
      \msqu} \right) \ln \left| 1 - { \msqu \over \mgl^2} \right| +
  \left( 1 \leftrightarrow 2 \right) \right] \,.
\label{G2bar} 
\eea
Notice that, in this scheme, $F^{\rm 2\ell}$ and $G^{\rm 2\ell}$
do not depend explicitly on $Q$.
We also stress that, in terms of our renormalized quantities $(\msqu,
\msqd, h_b, A_b)$, the corrections have a smooth heavy gluino
limit. In fact, in contrast with the case of the ${\cal O}(\at \as)$
corrections, the gluino decouples for $\mgl \rightarrow \infty$, since
$\mgl\,G^{\rm 2\ell} \rightarrow 0$ and $F^{\rm 2\ell}$ reduces to the
first line of Eq.~(\ref{G3bar}).

\section{Input parameters}

Phenomenological analyses of the MSSM parameter space should exploit
the experimental information on the bottom mass.  Instead of
expressing such information with the pole mass $M_b$, it is convenient
to use directly the running mass, in the SM and in the $\ov{\rm DR}$
scheme, evaluated at the reference scale $Q_0 = 175$~GeV. 
Following a procedure outlined in \cite{ellis},
we take as input the SM bottom mass in the $\ov{\rm MS}$
scheme, $m_b(m_b)^{\ov{\rm MS}}_{\rm SM} = 4.23 \pm 0.08$ GeV, as
determined from the $\Upsilon$ masses \cite{mbrun}; we evolve it up to
the scale $Q_0$ by means of suitable renormalization group equations
\cite{mbrge}; finally, we convert it to the $\ov{\rm DR}$ scheme. The
result, which accounts for the resummation of the universal large QCD
logarithms, is:
\be
\label{mbsmart}
\ov{m}_b \equiv m_b(Q_0)_{\rm SM}^{\ov{\rm DR}} = 
2.74 \pm 0.05 \; {\rm GeV} \, .
\ee
The relation between $\hat{h}_b \equiv 
h_b(Q_0)_{\rm MSSM}^{\ov{\rm DR}}$ and $\ov{m}_b$ of 
Eq.~(\ref{mbsmart}) is given by:
\be
\hat{h}_b \equiv h_b(Q_0)_{\rm MSSM}^{\ov{\rm DR}} 
=  {\ov{m}_b \sqrt{2} \over v_1} 
{1 + \delta_b \over \left| 1 +
\epsilon_b \right| } \, ,
\label{mbnoantri}
\ee
where 
\bea
\delta_b &  = & {\as \over 3 \pi} 
\left\{ {3 \over 2} - \ln \frac{\mgl^2}{Q_0^2}
\phantom{\left(\frac{2\,\mgl^2-\msqu}{\mgl^2 - \msqu}\right)} 
\right.\nn\\
&&\nn\\
&& \left. +\frac12 \left[
\frac{\msqu}{\mgl^2 - \msqu} \left( 1 - \left(
\frac{2\,\mgl^2-\msqu}{\mgl^2 - \msqu} - \frac{4 \,\mgl A_b}{\diff}
\right)  \ln \frac{\mgl^2}{\msqu} \right)
+ (1 \leftrightarrow 2)
\right] \right\} \, , 
\label{deltab}
\eea
and
\be
\epsilon_b  \;=\;  
-{2 \, \as \over 3 \, \pi} \,\frac{\mgl\,\mu\,\tb}{ \msqu-\msqd}\,\left[ 
\frac{\msqu}{\msqu - \mgl^2} \,\ln {\msqu \over \mgl^2} -
\frac{\msqd}{\msqd - \mgl^2} \,\ln {\msqd \over \mgl^2}
\right]\, .
\label{epsilonb}
\ee
The running parameter $\hat{h}_b$ is the appropriate input quantity to
be used with the $\ov{\rm DR}$ result presented in Section~2, while
the formulae obtained in Section~3 should be used with $h_b =
\hat{h}_b - \delta h_b$, as defined in that section, evaluating
Eq.~(\ref{dhb}) for $Q=Q_0$.

Notice that in Eq.~(\ref{mbnoantri}) the large ${\cal O} (\as)$
threshold corrections \cite{hrs} parametrized by $\epsilon_b$ have
been resummed to all orders as in \cite{resum}.  With the same
strategy, we can easily include the ${\cal O} (\at)$ threshold
corrections to the bottom mass, which are expected to generate the
largest two--loop ${\cal O}(\at \ab)$ corrections to the neutral Higgs
boson masses. It is sufficient to add to $\epsilon_b$ the analogous
quantity
\be
\label{epsbpri}
\epsilon_b^{\,\prime} \;= \; 
-{\at \over 4 \, \pi} \, \frac{A_t\,\mu\,\tb}{ \msqut-\msqdt}\,\left[ 
\frac{\msqut}{\msqut - \mu^2 }\,\ln {\msqut \over \mu^2} -
\frac{\msqdt}{\msqdt - \mu^2 }\,\ln {\msqdt \over \mu^2}
\right]\,,
\ee 
where the mixing between gauginos and higgsinos has been 
neglected, so that the masses of the higgsinos coincide with $\mu$.

For computing the two--loop ${\cal O}(\at\as + \ab\as)$ corrected
Higgs masses, as will be done in the numerical examples of the next
section, a suitable specification must be given for the parameters
entering the tree-level mass matrix and the one--loop ${\cal O}(\at+
\ab)$ corrections. In our effective potential approach, the tree-level
mass matrix is expressed in terms of the pole mass $\ma$ and of the
$\ov{\rm DR}$ parameter $\tb$, evaluated at the reference scale $Q_0$,
while the renormalization of the $Z$ boson mass (whose numerical value
we fix at $\mz = 91.187$ GeV) does not affect the ${\cal O}(\at\as +
\ab\as)$ corrections.  The parameters $v=246.218$ GeV and $\mu$ first
appear at the one--loop level and do not receive corrections at ${\cal
O}(\as)$. For the top--stop sector, we take as input the top pole
mass, conventionally fixed at $M_t = 175$ GeV, and the parameters
$(m_{Q,\tilde{t}}\,, m_U, A_t)$ that can be derived by rotating the
diagonal matrix of the On--Shell (OS) stop masses by the angle $\ths$,
defined as in \cite{dsz}. Concerning the sbottom sector, additional
care is required, because of our non--trivial definition of $h_b$,
Eq.~(\ref{mbnoantri}), and of the fact that, at ${\cal O}(\as)$, the
parameter $m_{Q,\tilde{b}}$ entering the sbottom mass matrix differs
from the corresponding stop parameter $m_{Q,\tilde{t}}$ by a finite
shift \cite{eberl}.  We start by computing the renormalized coupling
$h_b$ as given by Eqs.~(\ref{mbsmart})--(\ref{epsilonb}) and
(\ref{dhb}). Then we compute $m_{Q,\tilde{b}}$ following the
prescription of \cite{eberl}.  Finally, we use the parameters $h_b$
and $m_{Q,\tilde{b}}$ to compute the actual values of the OS sbottom
masses and mixing angle. The remaining input quantities, appearing
only in the two--loop corrections, are the gluino mass $\mgl$ and the
strong coupling constant, whose numerical value we fix at $\as(Q_0) =
0.108$.

\section{Numerical examples}

We are now ready for some numerical examples. To prepare the ground,
we study the variation of our renormalized $h_b$ with respect to other
parameters, keeping the reference bottom mass $\ov{m}_b$ fixed to the
central value of Eq.~(\ref{mbsmart}).

\begin{figure}[t]
\begin{center}
\mbox{
\hspace{-.4cm}
\epsfig{figure=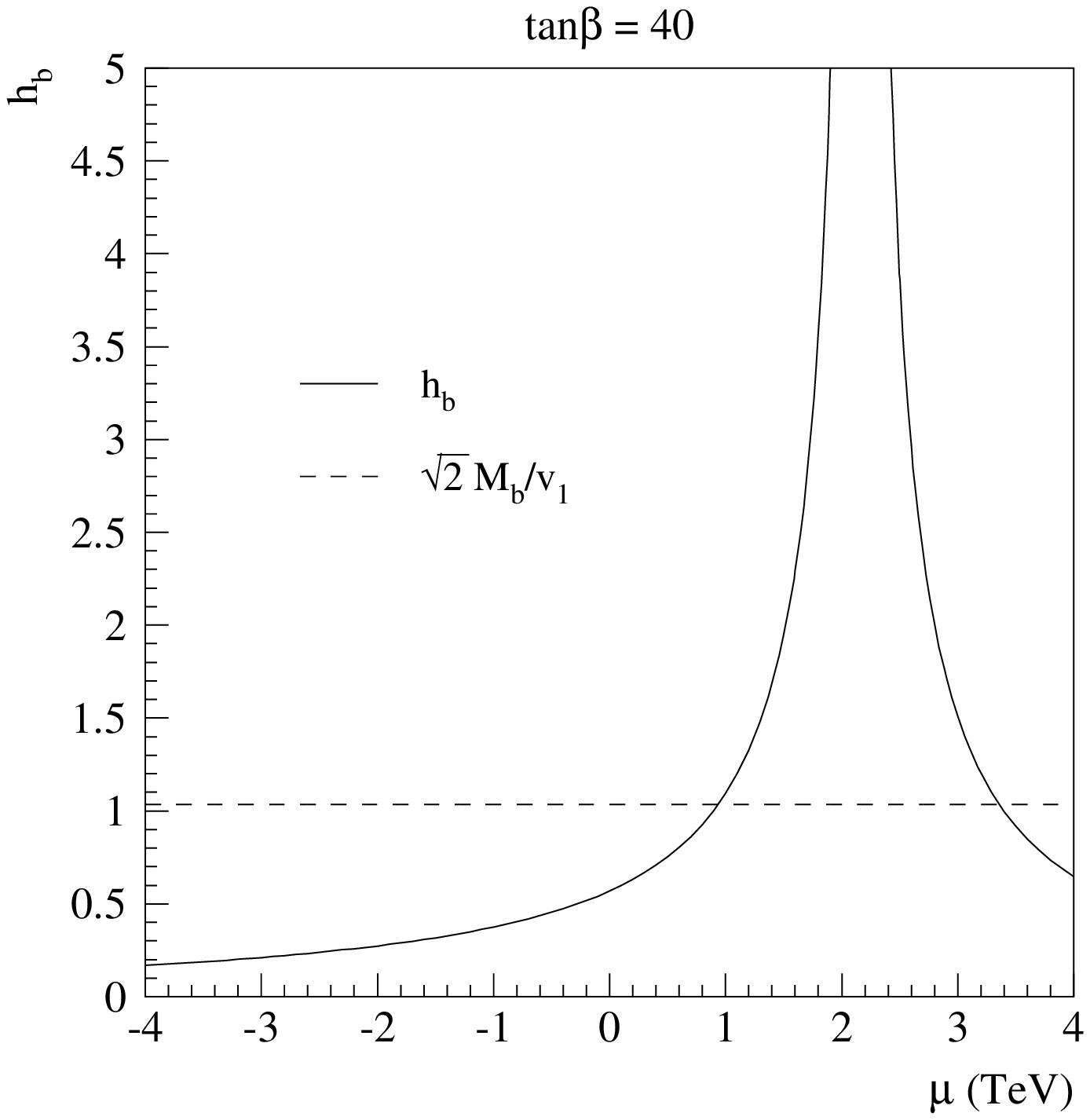,width=9cm,height=9cm}
\hspace{-1cm}
\epsfig{figure=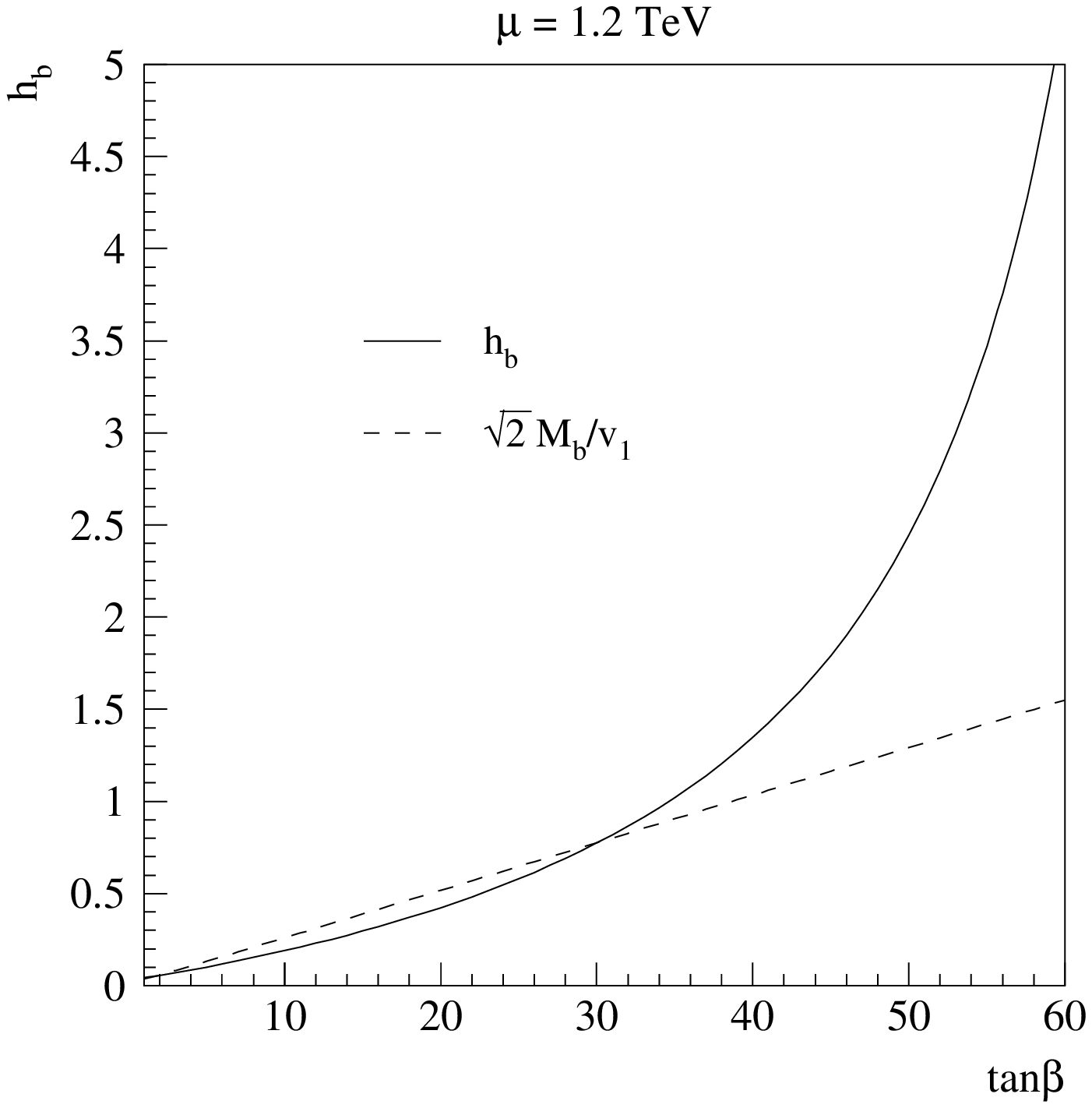,width=9cm,height=9cm}}
\end{center}
\vspace{-0.5cm}
\caption{The Yukawa coupling $h_b$, as defined in Eq.~(\ref{mbnoantri}): 
as a function of $\mu$ for
$\tan\beta = 40$ (left panel); as a function of $\tan\beta$ for
$\mu = 1.2$ TeV (right panel).  The other parameters are $A_b = 2$
TeV, $m_Q = m_D = \mgl = 1$ TeV. The quantity $h_b^{pole}
\equiv \sq2 \, M_b / v_1$ is also shown for comparison.}
\label{hbvsmutb}
\end{figure}

The left panel of Fig.~\ref{hbvsmutb} shows $h_b$ as a function of
$\mu$ (solid line), for $\tan\beta = 40$. The other relevant
parameters are chosen as $ A_b = 2$ TeV, $m_Q = m_D = \mgl = 1$ TeV
(the precise definition of $m_Q$ is not relevant in this case). The
quantity $h_b^{pole} = \sq2\,M_b/v_1$ is also shown as a dashed line.
The curve corresponding to $\hat{h}_b$ would be very close to that of 
$h_b$, thus we do not display it.  We see that having large values of 
$\tb$ and $\mu$ is a necessary but not sufficient condition for having 
a sizeable $h_b$: when the threshold contribution to the bottom mass
dominates, $|\epsilon_b| \gg 1$, $h_b$ must decrease for increasing
values of $|\mu| \tb$. We also see that, when there is an almost
complete destructive interference between the two contributions to
the bottom mass, $\epsilon_b \simeq -1$, the correct value of the
bottom mass cannot be reproduced by the one--loop formula for $h_b$ in
the perturbative regime, and the corresponding set of MSSM parameters
must be discarded.  Finally, we can see that the renormalized
$h_b$ can be large only for positive~\footnote{Our convention for the
sign of $\mu$ is implicitly defined in Eq.~(\ref{s2b}).} values of
$\mu$.  We then focus our attention on the case in which $\mu$ is
large and positive, so that $h_b$ and the corresponding corrections to
the Higgs masses can be sizeable.

For completeness, we should mention (for recent discussions and
references, see e.g. \cite{signmu}) that models with $b$--$\tau$
Yukawa coupling unification at the GUT scale favour, in our
conventions, a positive sign of $\mu \,\mgl$, which leads to a
negative $\epsilon_b$.  For sufficiently small $| \mu |$, radiative B
decays and the muon anomalous magnetic moment may favour a negative
sign of $\mu\, M_2$, where $M_2$ is the $SU(2)$ gaugino mass, and a
positive sign of $\mu\,A_t$. Similar but more model--dependent
constraints can be extracted, with the help of additional assumptions
on the soft supersymmetry--breaking terms, from the cosmological relic
density. Finally, having $\mu$ and $\tb$ simultaneously large may
require a certain amount of fine--tuning \cite{nr}.

The right panel of Fig.~\ref{hbvsmutb} shows $h_b$ as a function of
$\tan\beta$, for $\mu = 1.2$ TeV.  Again, the curve for $\hat{h}_b$
would be practically indistinguishable and we do not show it. The
other parameters are chosen as in the left panel, and the value of
$h_b^{pole}$ is also shown. We can see that, for this choice of
parameters (to be taken in the following as a representative one),
values of $\tan\beta$ much larger than 40--50 would imply a value of
$h_b$ beyond the perturbative regime. On the other hand, for low
values of $\tan\beta$ the coupling $h_b$ is even smaller than
$h_b^{pole}$, and the corresponding corrections to the Higgs masses
are expected to be negligible. For this reason, in the numerical
examples of the ${\cal O}(\ab\as)$ corrections we restrict ourselves
to values of $\tan\beta$ between 25 and 45.

\begin{figure}[p]
\begin{center}
\mbox{
\hspace{-.4cm}
\epsfig{figure=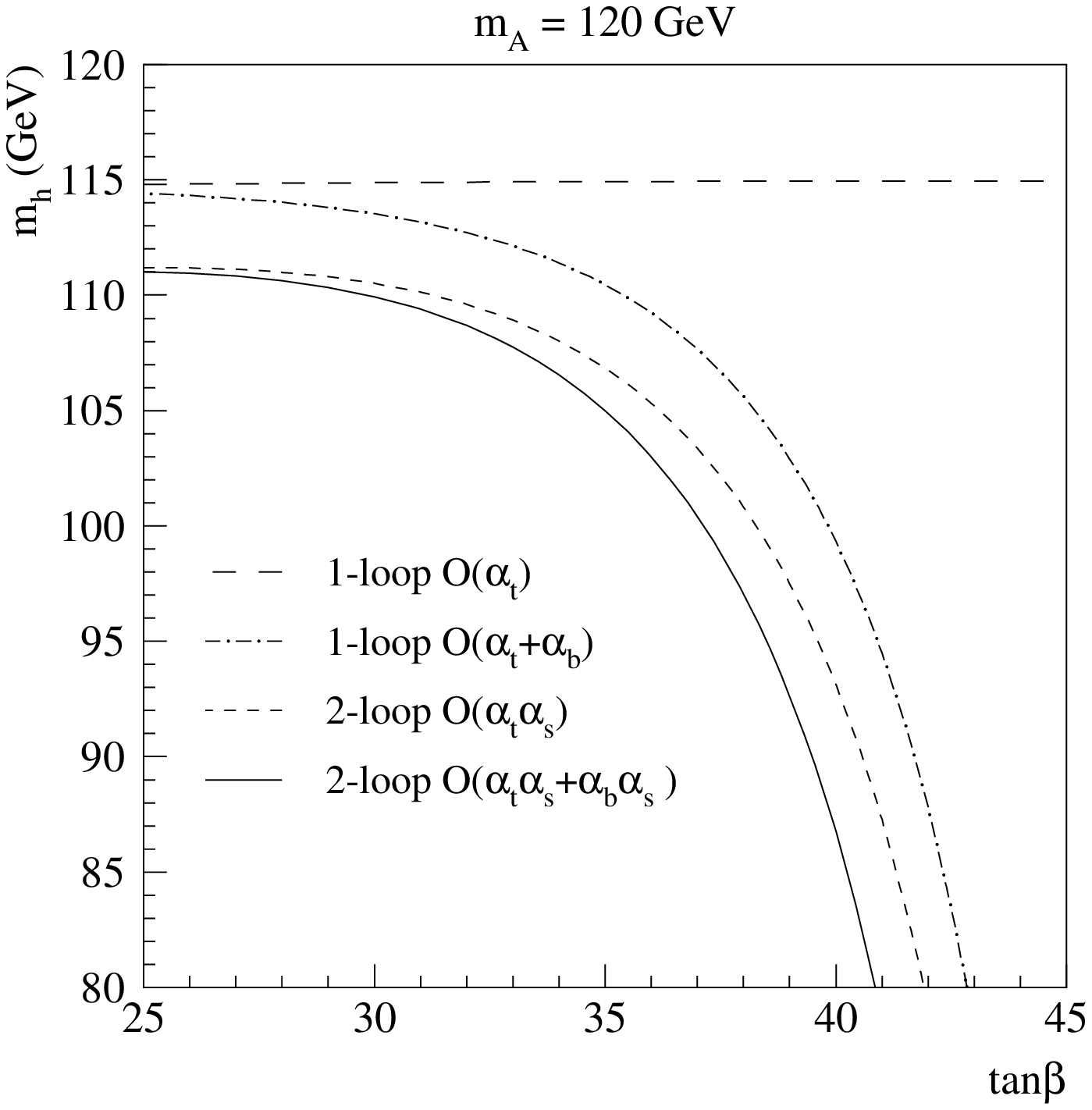,width=9cm,height=9cm}
\hspace{-1cm}
\epsfig{figure=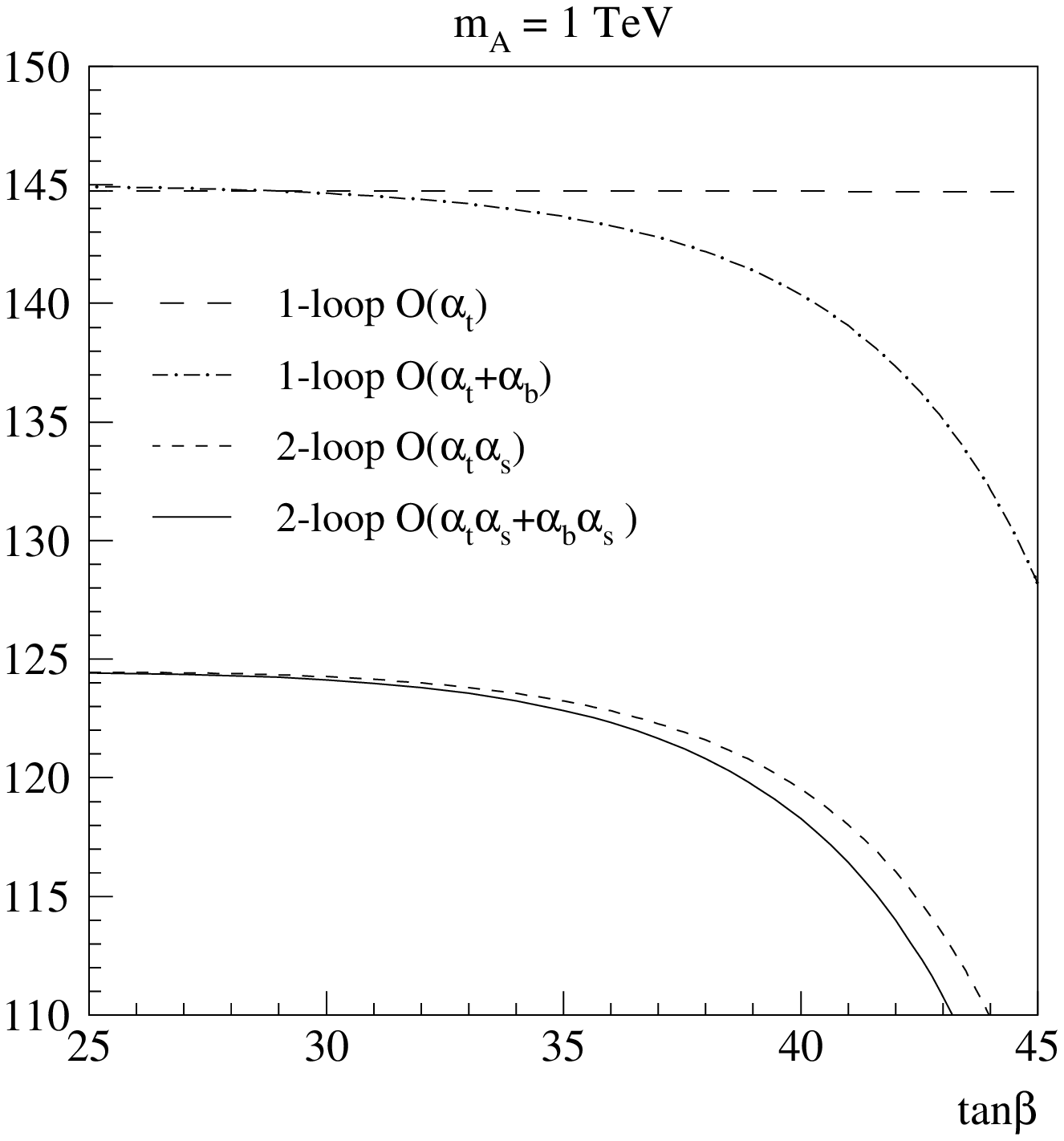,width=9cm,height=9cm}}
\end{center}
\vspace{-0.5cm}
\caption{The mass $m_h$ as a function of $\tan\beta$,
for $\ma = 120$ GeV (left panel) or 1 TeV (right panel).
The other parameters are $\mu = 1.2$ TeV, $A_t = A_b = 2$ TeV,
$m_{Q,\tilde{t}} = m_U = m_D = \mgl = 1$ TeV. 
The meaning of the different curves is explained in the text.}
\label{mhvstb}
\end{figure}
\begin{figure}[p]
\begin{center}
\mbox{
\hspace{-.4cm}
\epsfig{figure=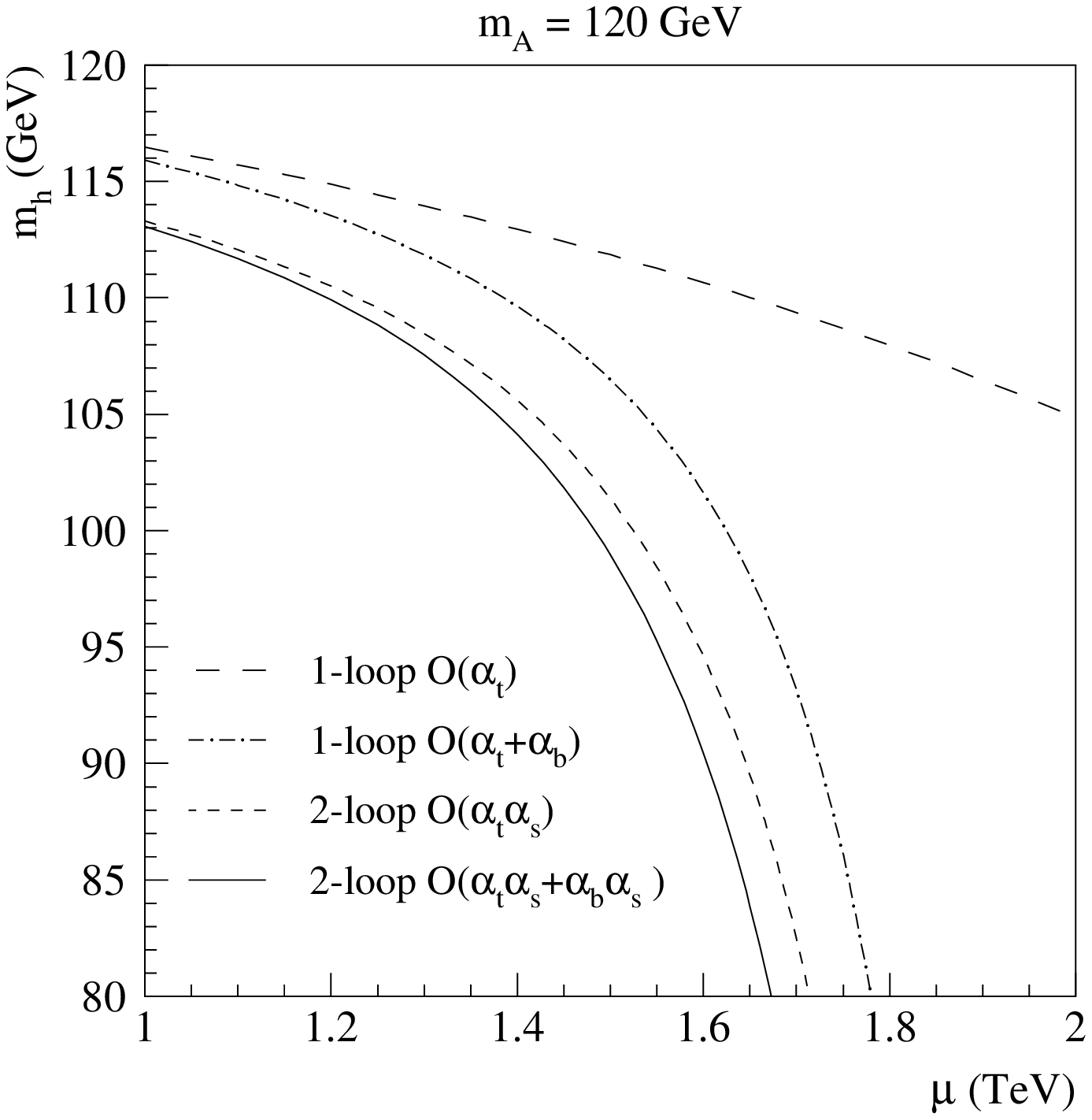,width=9cm,height=9cm}
\hspace{-1cm}
\epsfig{figure=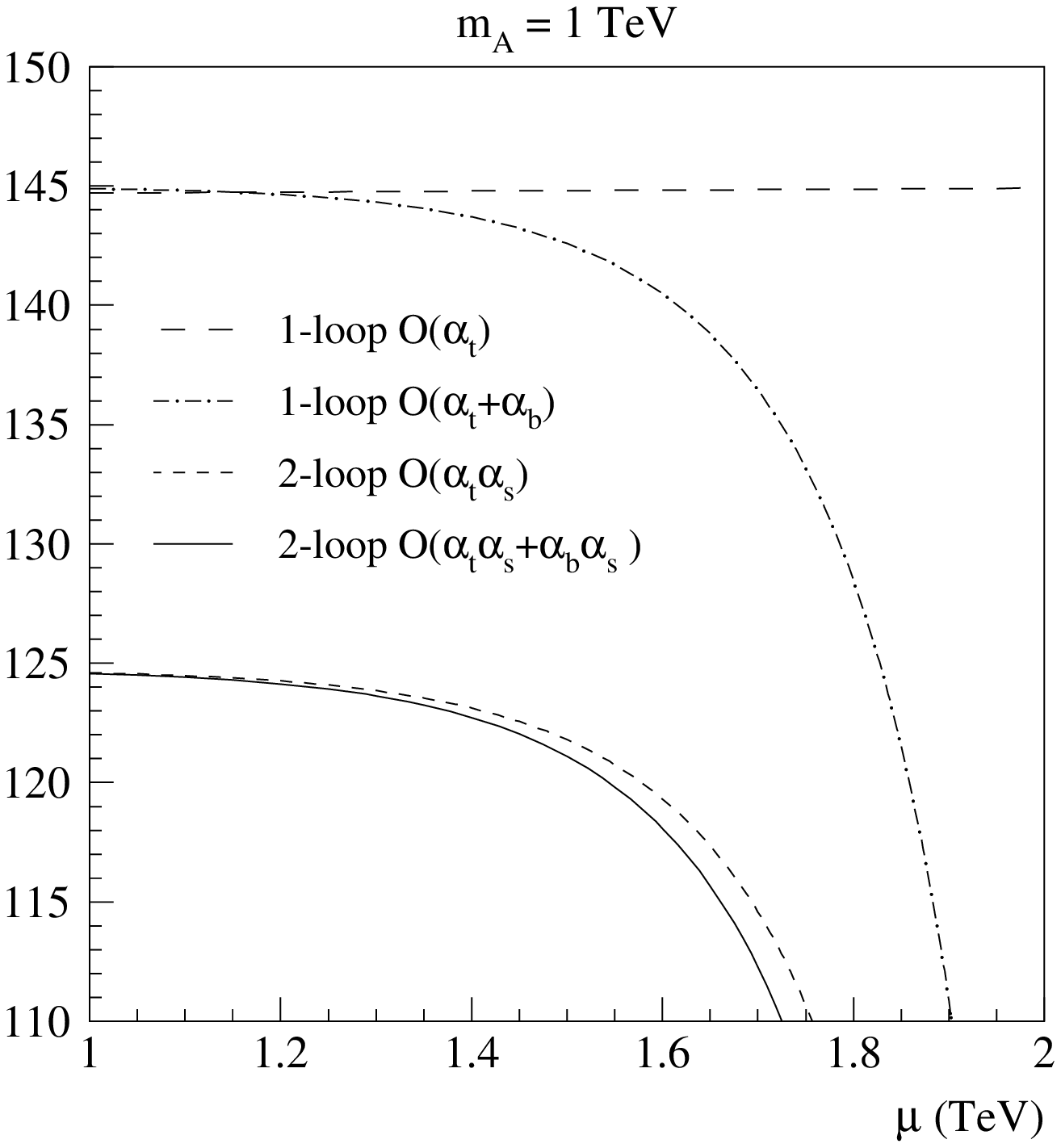,width=9cm,height=9cm}}
\end{center}
\vspace{-0.5cm}
\caption{The mass $m_h$ as a function of $\mu$,
for $\ma = 120$ GeV (left panel) or 1 TeV (right panel).
The other parameters are $\tan\beta = 30$, $A_t = A_b = 2$ TeV,
$m_{Q,\tilde{t}} = m_U = m_D = \mgl = 1$ TeV. 
The meaning of the different curves is explained in the text.}
\label{mhvsmu}
\end{figure}
Figs.~\ref{mhvstb} and \ref{mhvsmu} show the light Higgs mass $\mh$ as
a function of $\tan\beta$ for $\mu = 1.2$ TeV, and as a function of
$\mu$ for $\tan\beta = 30$, respectively.  In each figure, the left
panel corresponds to $\ma = 120$ GeV and the right panel to $\ma = 1$
TeV .  The other input parameters are chosen as $A_t = A_b = 2$ TeV,
$m_{Q,\tilde{t}} = m_U = m_D = \mgl = 1$ TeV. For this choice of
parameters, $m_{Q,\tilde{b}}$ differs from $m_{Q,\tilde{t}}$ by less
than 1\%. The curves in Figs.~\ref{mhvstb} and ~\ref{mhvsmu}
correspond to the one--loop corrected~\footnote{In the calculation of
the ${\cal O}(\at)$ and ${\cal O}(\ab)$ corrections we include the effects
proportional to $\mz^2$ and the momentum corrections as in \cite{ab}.} 
$\mh$ at ${\cal O}(\at)$ (long--dashed line) and at ${\cal O}(\at+\ab)$
(dot--dashed line), and to the two--loop corrected $\mh$ at ${\cal
O}(\at \as)$ (short--dashed line) and at ${\cal O}(\at\as+\ab\as)$
(solid line), respectively.  We can see from Fig.~\ref{mhvstb} that,
while the ${\cal O}(\at)$ prediction for $\mh$ is practically independent
of $\tan\beta$ for $\tan\beta>25$, the ${\cal O}(\ab)$ corrections
lower $m_h$ considerably when $\tan\beta$ increases. Fig.~\ref{mhvsmu}
shows that a similar decrease in $\mh$ occurs when $\mu$
increases. Both effects are enhanced by the steep dependence of the
renormalized coupling $h_b$ on $\tan\beta$ and $\mu$, depicted in
Fig.~\ref{hbvsmutb}. Comparing the solid and the short--dashed curves,
we can see that the `genuine' two--loop ${\cal O}(\ab\as)$ corrections
to the Higgs mass, given by Eqs.~(\ref{dmb11n})--(\ref{dmb22n}) and
(\ref{G3bar})--(\ref{G2bar}), are usually a small fraction of the
${\cal O}(\ab)$ ones, but the former can still reach several GeV when
the latter are very large. In particular, for small $\ma$ the ${\cal
O}(\ab\as)$ corrections can be comparable in magnitude with the ${\cal
O}(\at\as)$ ones. We stress that the absence of very large two--loop
effects from the sbottom sector is a consequence of our
renormalization prescription, which allows to set apart the
$\tan\beta$--enhanced corrections, resummed to all orders in the
renormalized coupling $h_b$. If we were to adopt for the sbottom
sector the same renormalization scheme that we use for the stop
sector, the dependence on $\tan\beta$ of the one--loop corrected $\mh$
would be smoother, but very large corrections (growing as
$\tan^2\beta$) would appear at two loops, questioning the validity of
the perturbative expansion.

\begin{figure}[t]
\begin{center}
\mbox{
\hspace{-.4cm}
\epsfig{figure=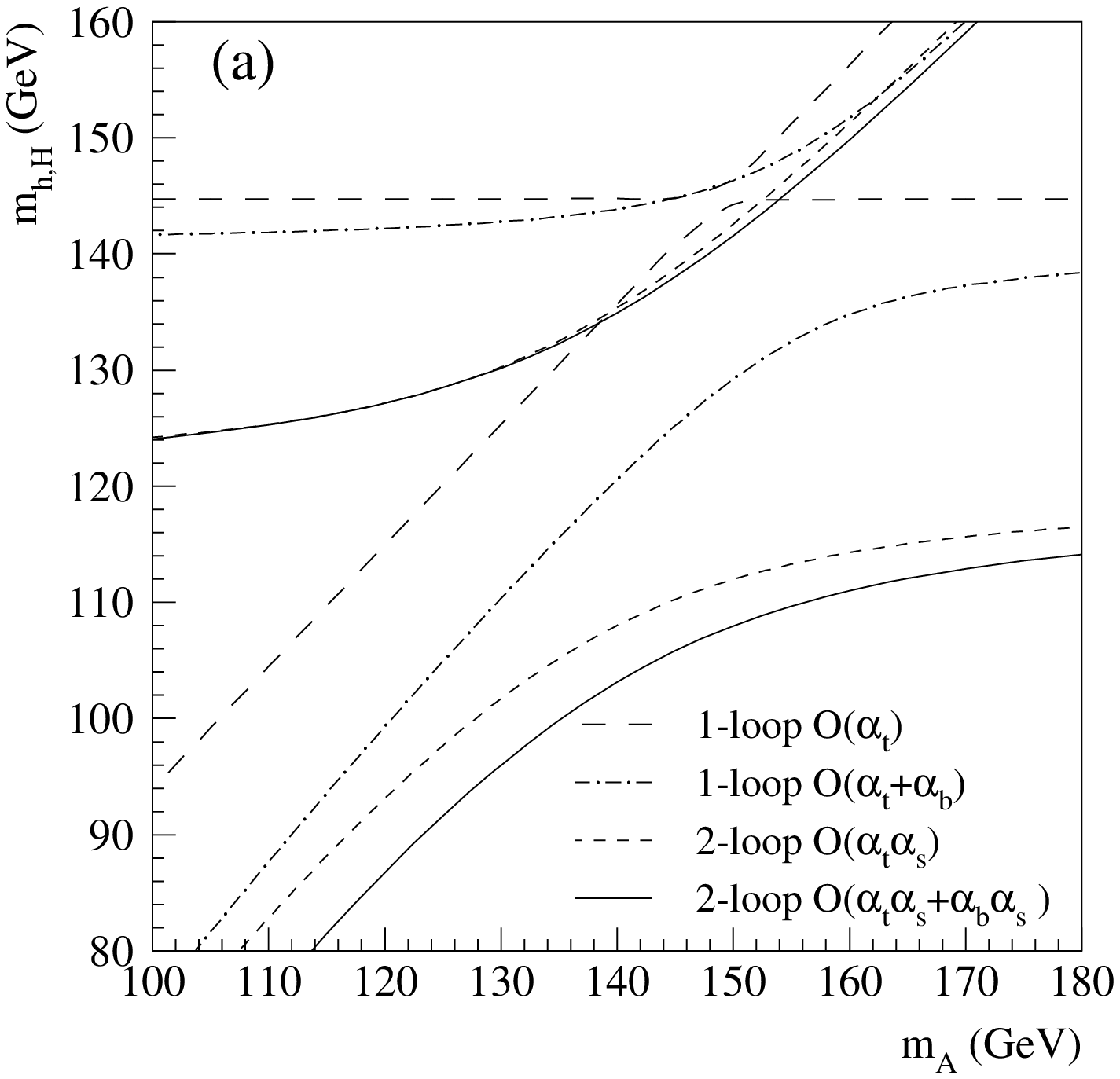,width=9cm,height=9cm}
\hspace{-1cm}
\epsfig{figure=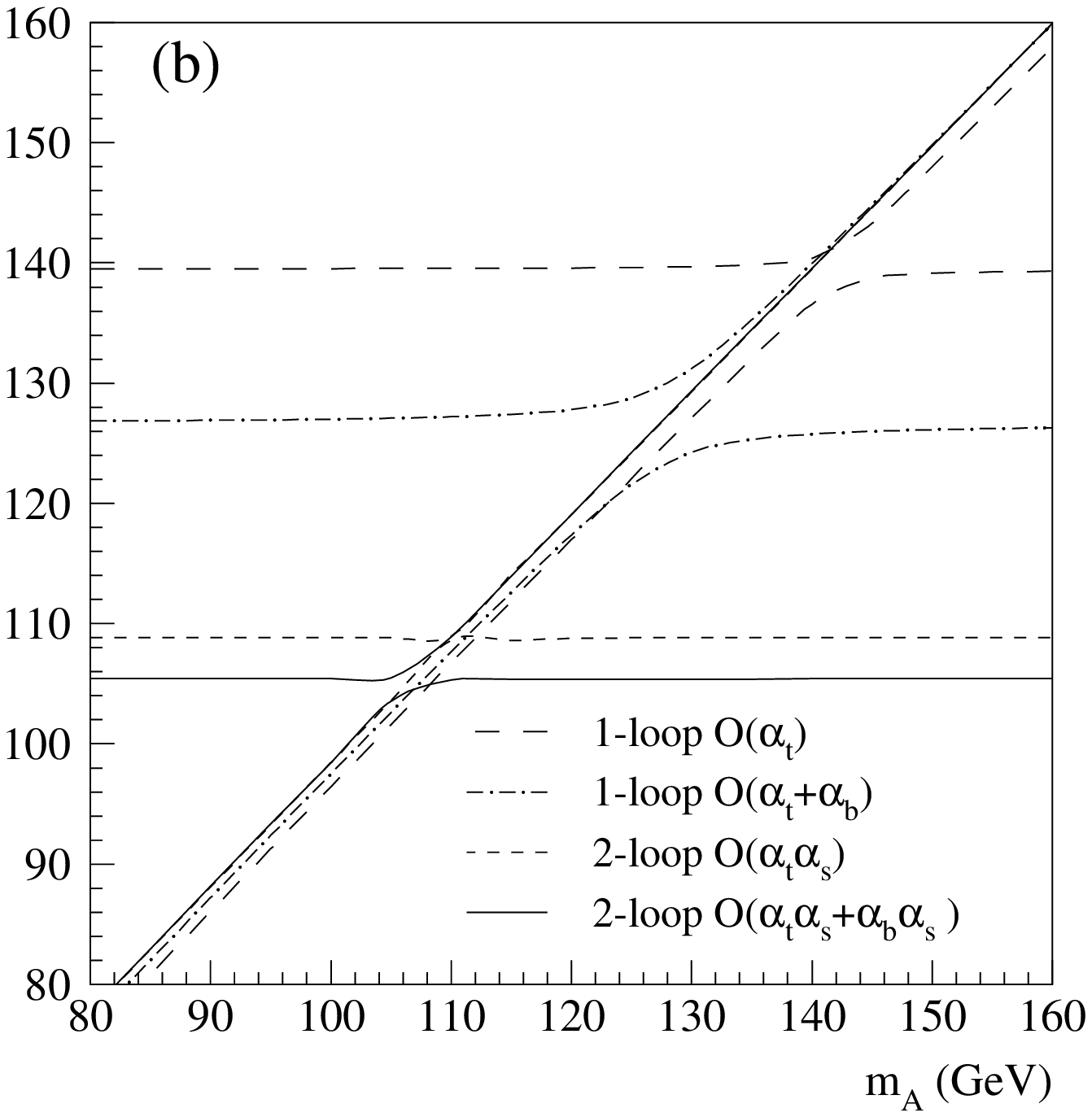,width=9cm,height=9cm}}
\end{center}
\vspace{-0.5cm}
\caption{The masses $m_h$ and $\mH$ as a function of $\ma$, for
$\mu = 1.2$ TeV and $m_{Q,\tilde{t}} = m_U = m_D = \mgl = 1$ TeV. 
The other parameters are (a) $\tan\beta = 40,\, A_t = A_b = 2$ TeV 
and (b) $\tan\beta = 45,\, A_t = 1.5$ TeV, $A_b = 0$. 
The meaning of the different curves is explained in the text.}
\label{mhhma}
\end{figure}
Finally, Figs. \ref{mhhma}a (left panel) and \ref{mhhma}b (right
panel) show both CP--even Higgs masses, $\mh$ and $\mH$, as functions
of the CP--odd Higgs mass, in the region of relatively small $\ma$ (80
GeV $< \ma <$ 180 GeV), for two different choices of the
parameters. In both figures we have chosen $\mu = 1.2$ TeV and
$m_{Q,\tilde{t}} = m_U = m_D = \mgl = 1$ TeV.  In Fig.~\ref{mhhma}a
the other parameters are $\tb = 40$ and $A_t = A_b = 2$ TeV.  From
Fig.~\ref{mhhma}a we see that, as anticipated above, when $\ma$ is
around 120 GeV the ${\cal O}(\ab\as)$ corrections to $\mh$ are of the
same size of the ${\cal O}(\at\as)$ ones. This is mainly due to the
large value of $A_b$, which enhances the correction to $(\Delta {\cal
M}^2_S )_{11}^{\rm eff}$, relevant for $\mh$ when $\ma$ is small.  The
${\cal O}(\ab\as)$ corrections to $\mH$ are rather small in this
example, but they can be larger for different parameter choices.  In
Fig.~\ref{mhhma}b the relevant parameters are $\tb = 45\,,\, A_t =
1.5$ TeV and $A_b= 0$.  For this choice, radiative corrections mainly
affect $(\Delta {\cal M}^2_S )_{22}^{\rm eff}$. Thus one of the
eigenvalues is roughly degenerate with $\ma$ and receives small
corrections, while the other eigenvalue is almost independent of $\ma$
and receives large corrections. In particular, the genuine ${\cal
O}(\ab\as)$ corrections to either $\mh$ or $\mH$ are around 3 GeV in
this example.

\section{Conclusions and discussion}

In this paper we presented explicit and general results for the ${\cal
O}(\ab \,\as)$ corrections to the MSSM neutral Higgs boson masses, in
the physically relevant limit of large $\tb$. Actually, a large value
of $\tb$ is a necessary but not a sufficient condition for having
large corrections, which require sizeable values of both $\mu$ and
$h_b$. We proposed a renormalization prescription for the sbottom
sector that automatically includes the decoupling of heavy gluinos and
separates the large threshold corrections, appearing in the relation
between $h_b$ and the pole bottom mass, from the genuine two--loop
effects. We also discussed the numerical impact of our results in a
number of representative examples.

A complete study of the two--loop (s)bottom corrections would require
also the knowledge of the ${\cal O}(\ab\at)$ and ${\cal O}(\ab^2)$
effects. Concerning the former, it is plausible that the most
important effects can be taken into account by adding to $\epsilon_b$
the analogous quantity $\epsilon_b^{\,\prime}$.  The ${\cal O}(\ab^2)$
corrections would need a dedicated calculation, but an estimate of
their importance can be obtained from our knowledge of the ${\cal
O}(\at^2)$ corrections. In Refs.~\cite{atsq,bdsz}, explicit formulae
for the ${\cal O}(\at^2)$ corrections to the Higgs masses, valid under
simplifying assumptions on the MSSM parameters, were presented. The
corresponding formulae for the ${\cal O}(\ab^2)$ corrections can be
derived from such formulae by performing suitable substitutions and
taking appropriate limits. In the case of large $\tan\beta$ and
universal soft sbottom masses, degenerate with $\ma$ and much larger
than the weak scale $(m_Q = m_D = \ma \equiv M_S \gg v)$, it is
possible to derive a simple expression for the ${\cal O}(\ab + \ab
\,\as + \ab^2)$ corrections to $\mh^2$:

\bea
\Delta m_h^2 & \simeq & -
\frac{\widetilde{X}_b^4}{8\,\pi^2\,M_S^4\,v^2}\,\left\{
1 + \frac{4\,\as}{3\,\pi}\,f \left(\frac{\mgl^2}{M_S^2}\right) 
+ \frac{3\,\ab}{4\,\pi}\,\left[\; 
f \left(\frac{\mu^2}{M_S^2}\right) -\frac{5}{2}
\right. \right. \nn\\
&&\nn\\
&& \left. \left.
- \frac{\mu^2}{M_S^2}\,\left( 2 \, \ln\frac{|\widetilde{X}_b|}{M_S^2}
+\frac{4}{3}\,\ln 2 + 1 \right)
+ C \, \frac{A_b^2}{M_S^2}
\right]\,\right\} \, ,
\label{absq}
\eea
where $\widetilde{X}_b = h_b\, v\,\mu/ \sqrt{2}$ in the large $\tb$
limit, $C \simeq 0.27$ and $f(x)$ is a positive function, defined as
\be
f(x)  \; = \;
\frac{x \, (3\,x-2)}{x-1}
- \frac{x^2\,( 3\,x^2 - 8\,x + 6)}{(x-1)^2}\,\ln x 
+ (3\,x + 1)\,(x-1) \,\ln |x-1|\; .
\ee
Some limiting values are $f(0) = 0\,,\; f(1) = 9/2, \:f(\infty) =
3/2$. In view of the result in Eq.~(\ref{absq}), we expect that, for
values of $\ab$ not much larger than $\as$, the ${\cal O}(\ab^2)$
corrections should be at most comparable with the `genuine' ${\cal
O}(\ab\,\as)$ effects.

%
\section*{Acknowledgments}
P.S. thanks A.~Dedes, A.~Quadt and V.~Spanos for discussions. F.Z. thanks
E.~Franco and G.~Martinelli for discussions, the Physics Department of
the University of Padua for its hospitality during part of this
project, and INFN, Sezione di Padova, for partial travel support. This
work was partially supported by the European Programmes
HPRN-CT-2000-00149 (Collider Physics) and HPRN-CT-2000-00148 (Across
the Energy Frontier).
%
%

%
\end{document}